\documentclass[useAMS,usenatbib]{mn2e}

\usepackage{graphicx}
\usepackage{color}
\usepackage{amssymb}

\title[Recurrent and symbiotic novae in the OGLE data]{Recurrent and symbiotic novae in the OGLE data}
\author[P. Mr\'oz et al.]{P. Mr\'oz$^{1}$\thanks{E-mail: pmroz@astrouw.edu.pl}, R. Poleski$^{1,2}$, A. Udalski$^{1}$, I. Soszy\'nski$^{1}$, M.~K. Szyma\'nski$^{1}$, M. Kubiak$^{1}$, \newauthor G. Pietrzy\'nski$^{1,3}$, \L{}. Wyrzykowski$^{1,4}$, K. Ulaczyk$^{1}$, S. Koz\l{}owski$^{1}$, P. Pietrukowicz$^{1}$, \newauthor and J. Skowron$^{1}$\\
$^{1}$Warsaw University Observatory, Al. Ujazdowskie 4, 00-478 Warszawa, Poland\\
$^{2}$Department of Astronomy, Ohio State University, 140 W. 18th Ave., Columbus, OH 43210, USA\\
$^{3}$Universidad de Concepci\'on, Departamento de Astronom\'ia, Casilla 160-C, Concepci\'on, Chile\\
$^{4}$Institute of Astronomy, University of Cambridge, Madingley Road, Cambridge CB3 0HA, UK\\}
\begin{document}

\date{Accepted 2014 June 12. Received 2014 June 12; in original form 2014 May 07}

\pagerange{\pageref{firstpage}--\pageref{lastpage}} \pubyear{2014}

\maketitle

\label{firstpage}

\begin{abstract}
We analyse long-term optical photometry for two Galactic recurrent novae (V745 Sco and V3890 Sgr) and one Large Magellanic Cloud object (Nova LMC 1990b) observed over several years by the Optical Gravitational Lensing Experiment (OGLE) sky survey. We do not find variability with the previously claimed orbital period of V745 Sco. This voids previous findings based on this value (e.g. the distance determination). The quiescence variability of this object is dominated by semiregular pulsations of the red giant secondary (with periods of 136.5 and 77.4 d). The photometry of Nova LMC 1990b reveals an unnoticed eruption in 2010 and eclipse-like variability in quiescence with a period of 1.26432(8) d. The photometric properties make this object very similar to U Sco. Finally, we describe the eruptions of two likely symbiotic novae, V5590 Sgr and OGLE-2011-BLG-1444. The secondary of V5590~Sgr is a Mira star with pulsation period of 236 d.
\end{abstract}


\section{Introduction}

A classical nova eruption is caused by a thermonuclear runaway on the surface of a white dwarf in a binary system (Bode \& Evans 2008). By definition, any star that shows a second such explosion becomes a recurrent nova (RN). From a theoretical point of view, two conditions must be fulfilled to obtain a short inter-eruption time-scale: the mass-transfer rate must be high ($\gtrsim 10^{-7} M_{\odot}\ \rm{yr}^{-1}$) and the mass of the white dwarf has to be larger than 1.2~$M_{\odot}$ (e.g. Kato et al. 2014).

For this reason, some RNe are suspected to be progenitors of Type Ia supernovae. However, their estimated number is too low to account for the observed supernova rate (e.g. Maoz, Mannucci \& Nelemans 2014, and references therein).

There are only 10 RNe known in the Milky Way (i.e. T Pyx, IM Nor, CI Aql, V2487 Oph, U Sco, V394 CrA, T~CrB, RS Oph, V745 Sco and V3890 Sgr), three in the Large Magellanic Cloud (LMC 1990b, YY Dor, and, most likely, LMC 2009; Shafter 2013) and a few in M31 (including the remarkable M31N 2008-12a,  which has shown five nova eruptions in five years; Darnley et al. 2014; Tang et al. 2014), but many explosions could have been missed. For an exhaustive review of galactic RNe, we refer the reader to \citet{sch}.

Another class of interesting objects is that of symbiotic novae (SyNe) with evolved red giant secondaries. Some of these might be recurrent (i.e. T CrB, RS Oph, V745 Sco, V3890~Sgr). However, the term 'symbiotic novae' is usually restricted to objects with prolonged eruptions (lasting from years to decades), when steady hydrogen burning at the white dwarf surface occurs. There are only a few SyNe known (e.g. AG Peg, PU Vul, RR Tel, HM Sge). A recent review has been given by Miko\l{}ajewska (2010).

Regular long-term observations of dense stellar regions of the sky by the Optical Gravitational Lensing Experiment (OGLE) photometric survey have allowed unprecedented studies of very rare stellar events, such as the detection of the binary star merger V1309 Sco \citep{tyl}, or the discovery of mode switching in the RR Lyrae star OGLE-BLG-RRLYR-12245 \citep{sos14}. In this paper, we present an analysis of the photometric properties of three RNe and two SyNe observed by the OGLE project. 

\begin{table*}
\centering
\begin{tabular}{lccccccc}
\hline
\multicolumn{1}{c}{Name} & RA & Dec. & Field & Star ID & Field & Star ID \\ 
     & J2000.0 & J2000.0 & (OGLE-III) & (OGLE-III) & (OGLE-IV) & (OGLE-IV) \\ \hline
Nova LMC 1990b     & 05$^{\rm{h}}$09$^{\rm{m}}$58\fs40 & -71$^{\circ}$39${'}$52\farcs7 & --       & --     & LMC508.01 & 1937  \\
OGLE-2011-BLG-1444 & 17$^{\rm{h}}$50$^{\rm{m}}$19\fs26 & -33$^{\circ}$39${'}$07\farcs2 & BLG140.7 & 169918 & BLG502.14 & 59301 \\
V745 Sco           & 17$^{\rm{h}}$55$^{\rm{m}}$22\fs22 & -33$^{\circ}$14${'}$58\farcs6 & BLG149.1 & 78644  & BLG508.07 & 67742 \\
V5590 Sgr          & 18$^{\rm{h}}$11$^{\rm{m}}$03\fs70 & -27$^{\circ}$17${'}$29\farcs4 & BLG236.1 & --     & BLG523.25 & --    \\
V3890 Sgr          & 18$^{\rm{h}}$30$^{\rm{m}}$43\fs28 & -24$^{\circ}$01${'}$08\farcs9 & BLG278.6 & 85802  & --        & --    \\
\hline
\end{tabular} 
\caption{Basic data of the described objects.}
\label{basicdata}
\end{table*}

\section[]{Observations and Data Reductions}

The data presented in this paper were collected during the third and fourth phases of the OGLE (OGLE-III, OGLE-IV; Udalski 2003) in 2001--2009 and 2010--2013, respectively. The observations were conducted with the 1.3-m Warsaw Telescope located at Las Campanas Observatory, which is operated by the Carnegie Institution for Science. 

The measurements were taken through the $I$- and $V$-band filters, closely resembling those of the standard Johnson-Cousins system. The photometry was carried out with the Difference Image Analysis (DIA) algorithm (Alard \& Lupton 1998; Wo\'zniak 2000). Details of the instrumentation, reductions, astrometric and photometric calibrations can be found in \citet{uda}, \citet{uda3} and \citet{szy}.

In table \ref{basicdata}, we give basic information for the described objects.

The time-series $I$- and $V$-band photometry is available to the astronomical community from the OGLE Internet Archive.\footnote{See http://ogle.astrouw.edu.pl and ftp://ftp.astrouw.edu.pl/ogle/ogle4/NOVAE.}

\begin{figure*}
 \includegraphics[width=\textwidth]{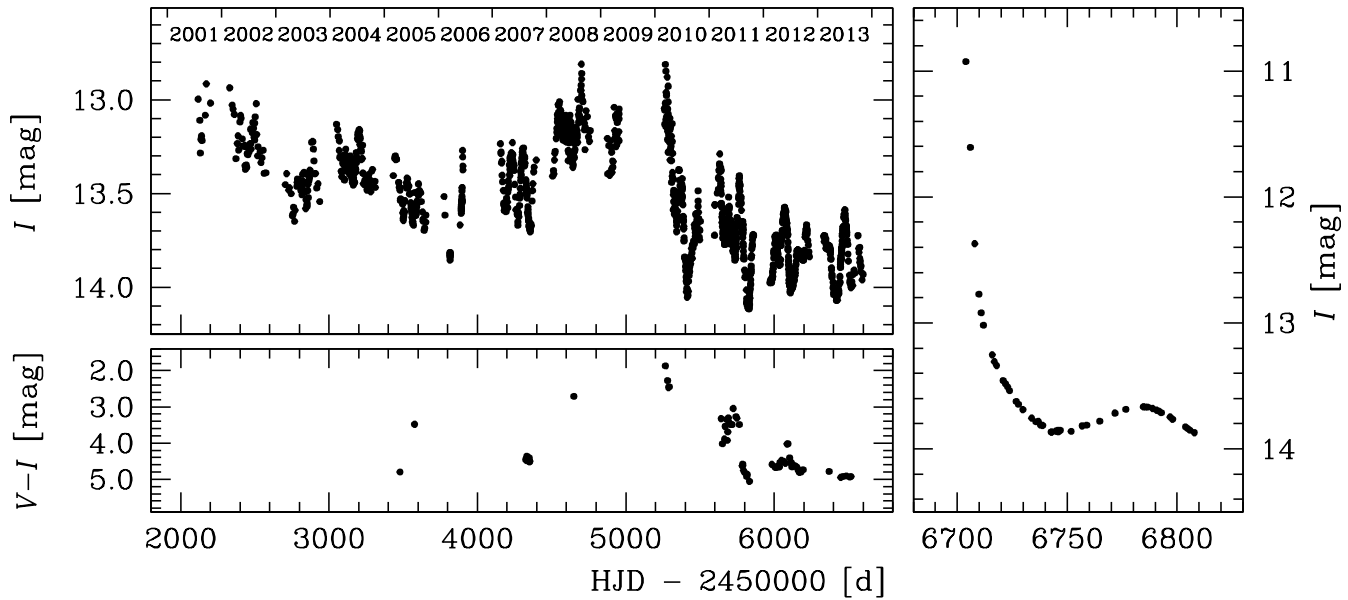}
 \caption{OGLE light and colour curves of V745 Sco. The right panel shows a close-up of the 2014 eruption.}
 \label{fig1}
\end{figure*}

\section[]{V745 Scorpii}

Because of its location towards the dense Galactic bulge regions, V745 Sco is a relatively poorly studied RN. Its eruptions were observed in 1937 (Plaut 1958), 1989 (Liller 1989; Sekiguchi et al. 1990b) and, most recently, in 2014 February (Stubbings 2014). The latest observed eruption has enabled intensive multiwavelength studies of this object (e.g. Banerjee et al. 2014, and references therein). The time intervals between eruptions are 52.2 and 24.5 yr with a possible average recurrence time-scale $\tau_{\rm{rec}}\approx 25.5$ yr. The secondary is a red giant or supergiant of M type (Harrison et al. 1993; Anupama \& Miko\l{}ajewska 1999). \citet{sch09} measured the orbital period $P=510 \pm 20$ d and, on that basis, \citet{sch} calculated the distance, spectral energy distribution and other properties. However, our observations do not confirm his result.

The $I$-band light curve of V745 Sco consists of 1581 measurements from 1269 nights spanning the period 2001--2014. The nova is located in the OGLE-IV field BLG508, which is observed one to three times per night. Unfortunately, the beginning of the 2014 eruption was not observed because of the proximity of the star to the Sun in the sky. The star was already overexposed in the first images taken in 2014 February. The entire light curve is shown in Fig. \ref{fig1}.

Additionally, 83 $V$-band measurements (from 69 nights) were collected. Based on the closest $I$-band point, we estimated the $V-I$ colour (see the lower-left panel of Fig.~\ref{fig1}). The median time interval between $V$- and $I$-band images is 1.1 h. Keeping in mind fast (1--1.5 h) flickering \citep{sch}, the accuracy of the colour measurements is limited to 0.05--0.1 mag.

\subsection[]{Orbital period}

The most prominent features of the quiescent light curve are red giant pulsations superimposed on long-term variations. In fact, V745 Sco was included in the OGLE-III Catalogue of Variable Stars as OGLE-BLG-LPV-120729 \citep{sos} and classified as a semiregular variable (SRV). \citet{sos} reported pulsation periods of 78.63 and 137.53~d, and $I$-band amplitudes of 0.193 and 0.158~mag, typical for this class of variable stars.

\begin{figure}
 \includegraphics[width=0.45\textwidth]{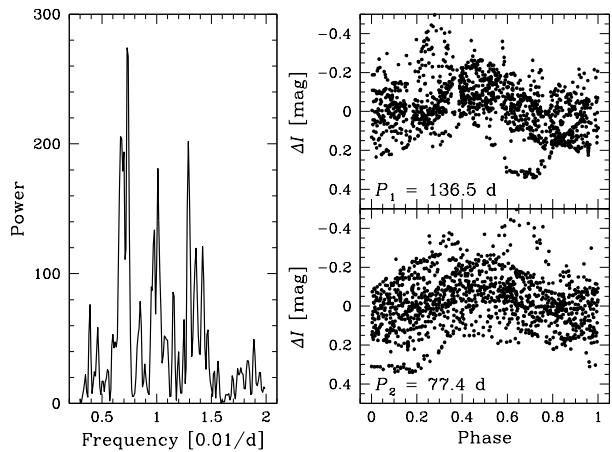}
 \caption{Red giant pulsations in V745 Sco. Left panel: the Lomb-Scargle periodogram. Right panel: the quiescent light curve folded with pulsation periods $P_1 = 136.5$ d and $P_2 = 77.4$~d.}
 \label{fig2}
\end{figure}

In order to measure the pulsation period, we changed magnitudes into flux and detrended the light curve. Then, we used the Lomb-Scargle algorithm \citep{sca} and refined periods with the {\sc tatry} code \citep{shw}, which uses the multiharmonic analysis of variance. The periodogram (see the left panel of Fig. \ref{fig2}) has two strong peaks at $P_1 = 136.5 \pm 1.0$~d and $P_2 = 77.4 \pm 1.0$ d, which likely correspond to the fundamental-mode and first-overtone radial pulsations (e.g. Soszy\'nski, Wood \& Udalski 2013), respectively. The third peak, located at~$f = 1/P_1 + 1/365 \approx 0.01$~d$^{-1}$, is a year alias of $P_1$. After pre-whitenings, the remaining peaks are much weaker and they are associated with a variability that has about an order of magnitude lower amplitude. The right panel of Fig. \ref{fig2} shows the light curve folded with periods $P_1$ (upper-right plot) and $P_2$ (lower-right plot).

We fitted sinusoids corresponding to $P_1$ and $P_2$ and subtracted them from the original data. We searched such pre-whitened data for additional periodicity in a range 10--10000 d. The most prominent peaks correspond to $P_3 = 2413 \pm 6$ d, $P_4 = 344.1 \pm 0.7$~d, $P_5 = 199.5 \pm 0.5$~d and $P_6 = 159.5 \pm 0.2$ d. We cannot exclude some orbital-related variability as a cause, but detected periods are probably produced by long-term variations and/or pulsations.

\begin{figure}
 \includegraphics[width=0.45\textwidth]{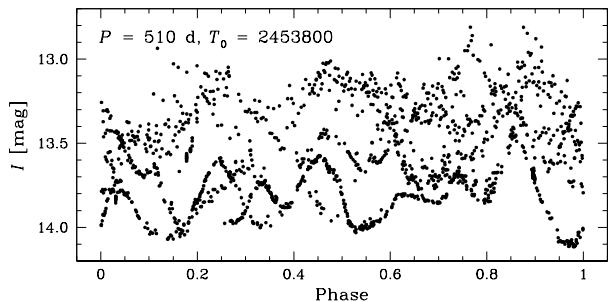}
 \caption{The light curve of V745 Sco folded with the ephemeris from \citet{sch09}. No ellipsoidal effect is visible.}
 \label{fig6}
\end{figure}

Schaefer (2009) used data from 246 nights, collected between 2004 June and 2008 August, and measured a photometric periodicity of $255 \pm 10$ d (and attributed it to ellipsoidal modulations). In our data, limited to the time-span of the observations of Schaefer (2009), we detected periods of 136.6 and 77.9 d. There was no statistically significant periodicity at $255 \pm 10$ d (see Fig. \ref{fig2}). We also phased our light curve with the ephemeris given in \citet{sch09}, but clearly no ellipsoidal effect is visible (Fig. \ref{fig6}). This signal should definitely be strong in the $I$-band, where the red giant contribution dominates.

We suspect that the inclination of this system is low and that ellipsoidal variations are either of small amplitude, compared to pulsations, flickering and long-term variability, or simply invisible. The amplitudes of ellipsoidal variables are smaller than $\sim$0.3 mag; it is almost impossible to distinguish such changes from large-amplitude, irregular variations occuring on the same time-scale. Therefore, the orbital period can be detected only spectroscopically (which is not easy in this case). Because the secondary is a red giant (or supergiant), we are left with a crude approximation of the orbital period of a few hundreds of days (in other RNe with red giant secondaries, for example, 228~d for T~CrB and 457~d for RS~Oph; Schaefer 2010).

\subsection[]{Pre-eruption light curve}

The OGLE light curve of V745 Sco enables studies of long-term brightness variations before the 2014 eruption. Between 2006 and 2008, V745 Sco brightened at a rate of 0.15 mag/yr. In 2009--2010, it reached a maximum at $I \sim 12.8$ mag followed by a fast decline by 1 mag in 150 days. Subsequently, the mean brightness decreased monotonically until 2013 November. In the $V$ band, the star faded from $\sim$15 mag to 19 mag, throughout 2010 and 2011.

The observed brightness of the object in quiescence is composed of the contributions from the accretion disk, hotspot, and red giant. The $V$-band peak-to-peak amplitude of the pulsating red giant is 1.0 -- 1.2 mag, so it cannot be fully responsible for the described drop. The observed flux decreased by a factor of at least 20, which suggests a similar decline for the accretion rate. Similar flares (from months to 2 yr in duration and amplitudes up to 2 mag) were observed in RS~Oph (Schaefer 2010). It is not clear if such behaviour was connected with the 2014 eruption. In analogy to RS~Oph, where flares are seen at random times and are not associated with explosions, we can rule this out. However, Adamakis et al. (2011) found some pre-eruption signal via wavelet analysis of the light curve of RS Oph.

For a few novae, a pre-eruption rise has been observed. The brightening started from 2 yr (V533 Her) to several days (T Pyx, V1500 Cyg) before the explosion (Collazzi et al. 2009; Schaefer et al. 2013). We can rule out the early pre-eruption rise, starting at least 90 d prior to the 2014 eruption. Because of a conjunction with the Sun, the photometric behaviour of the star between 2013 November and 2014 February will remain a mystery.

\subsection{2014 eruption}

The explosion was detected on 2014 February 6.7 UT (HJD = $2456695.2$) and reported by Stubbings (2014). The star was overexposed in the image taken on 2014 February 7.4 UT -- our photometry starts 8 d after the peak. The right panel of Fig. \ref{fig1} shows a close-up of the light variations during the eruption. A rate of decline decreased continuously from 0.44 to 0.01 mag/d. In 2014 March, $\sim$50 days after the peak, the nova returned to quiescence.  

The light curve is smooth during the decline -- an S type in the classification system of Strope et al. (2010) -- and we have not observed a plateau. The photometry from the American Association of Variable Star Observers (AAVSO) data base\footnote{http://www.aavso.org/data/lcg} also shows a fast, almost smooth, early decline from maximum in the $B$, $V$, and $R$ bands. Plateaux were found only in RNe (Schaefer 2010). However, some RNe with giant secondaries -- T CrB (Schaefer 2010) and apparently V745~Sco -- do not show this feature.

\section[]{V3890 Sagittari}

V3890 Sgr was observed in eruption in 1962 (Dinerstein \& Hoffleit 1973) and 1990 (Kilmartin et al. 1990). \citet{sch} reports an orbital period of 519.7~d and a distance of $7.0 \pm 1.6$~kpc. The secondary is an M5 red giant \citep{har}.

\begin{figure}
 \includegraphics[width=0.45\textwidth]{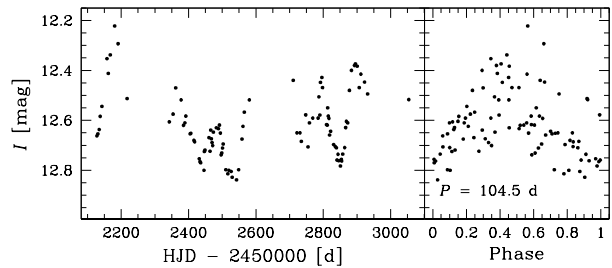}
 \caption{OGLE-III light curve of V3890 Sgr. The right panel presents the light curve phased with the pulsation period $P=104.5$~d.}
 \label{fig4}
\end{figure}

The OGLE $I$-band light curve of V3890 Sgr (shown in Fig. \ref{fig4}) consists of only 100 measurements spanning the period 2001--2003. The variability is dominated by the red giant pulsations with a period of $104.5 \pm 1.0$ d (the phased light curve is shown in the right panel of Fig. \ref{fig4}). Pulsations are superimposed on the long-term trend with an amplitude of $\sim 0.4$ mag. There are some weaker periodicities at $178.7 \pm 1.5$ d, $78.9 \pm 0.5$ d and $140.8 \pm 1.5$ d.

We have not detected photometric variations  with half the orbital period described by \citet{sch09} and \citet{sch}, because their amplitude decreases toward infrared (from 1.0 mag in $B$ to 0.07 mag in $J$). Moreover, in the $I$-band, any such signal is buried in red giant semiregular pulsations. Also, the span of observations is too short for sound detection of such a long periodicity.

\section[]{Nova LMC 1990b}

Nova LMC 1990b was the first extragalactic RN to be recognized (Sekiguchi et al. 1990a; Shore et al. 1991). Known eruptions occurred in 1968 December (Sievers 1970) and 1990 February. 

\subsection{2010 eruption}

The nova has been monitored in the OGLE-IV project since 2010. Our data set consists of 587 measurements in the $I$ band and 114 in the $V$ band. The light curve is shown in Fig. \ref{fig3}.

In 2010 November, Nova LMC 1990b underwent a third known eruption, unfortunately missed by the astronomical community. Because the real-time data reduction pipeline was not available at that time, we could not have alerted this event. The explosion has not been noticed by Walter et al. (2012) in their Stony Brook / SMARTS Atlas of (mostly) Southern Novae, although they have monitored old novae.

In the image taken on 2010 November 21.2 UT, the star was already overexposed. In the previous frame, taken on 2010 November 19.2 UT, the nova had a normal quiescent brightness around 19 mag, so the eruption and the maximum light must have occurred around 2010 November $20.2 \pm 1.0$ UT (HJD = $2455520.7 \pm 1.0$).

During the first week after the eruption, the nova faded at a rate of $0.57 \pm 0.07$ mag/d in $I$ and $0.59 \pm 0.05$ mag/d in $V$. We assessed the peak magnitudes $I_{\rm{peak}} = 11.4 \pm 0.4$ mag and $V_{\rm{peak}} = 11.7 \pm 0.3$ mag and estimated the decline time-scales $t_2 \approx 3.5$ d and $t_3 \approx 5$ d, respectively. 

Between 2010 November 26 and December 8, the nova showed a plateau. The brightness dropped slightly by $\sim0.5$ mag. Subsequently, the system faded slowly at a constant rate of $0.070 \pm 0.006$ mag/d in $I$ and $0.094 \pm 0.008$ mag/d in $V$. Two months after the eruption, it reached its normal quiescent brightness $I_{\rm{qui}}=19.2$ mag. Following the classification system of Strope et al. (2010), the nova is of P(5) type. 

The mean colour in quiescence is $\langle V-I \rangle_{\rm{qui}} = 0.37 \pm 0.05$ mag, while during the eruption the nova was slightly bluer, $\langle V-I \rangle_{\rm{erup}} = 0.29 \pm 0.03$ mag. The colour is nearly constant throughout the light curve, which is a typical feature of RNe.

Adopting the distance modulus $m-M = 18.49$ mag for the Large Magellanic Cloud (Pietrzy\'nski et al. 2013) and the mean extinction $A_V = 0.70 \pm 0.43$ mag (Zaritsky et al. 2004), we estimate absolute magnitudes at maximum brightness $M_{\rm{peak}}^V=-7.5$ mag and in quiescence $M_{\rm{qui}}^V = 0.4$~mag.

The inter-eruption intervals are 21.2 and 20.8 yr, so the average recurrence time-scale is $\tau_{\rm{rec}}\approx 21$~yr.

\subsection{Orbital period}

There is only one mention in the literature of the orbital period of Nova LMC 1990b, which is $P_{\rm{orb}} \sim 1.3$ d \citep{sek92}. We have used the {\sc tatry} code to find any periodic variations in the quiescence.

\begin{figure*}
 \includegraphics[width=0.72\textwidth]{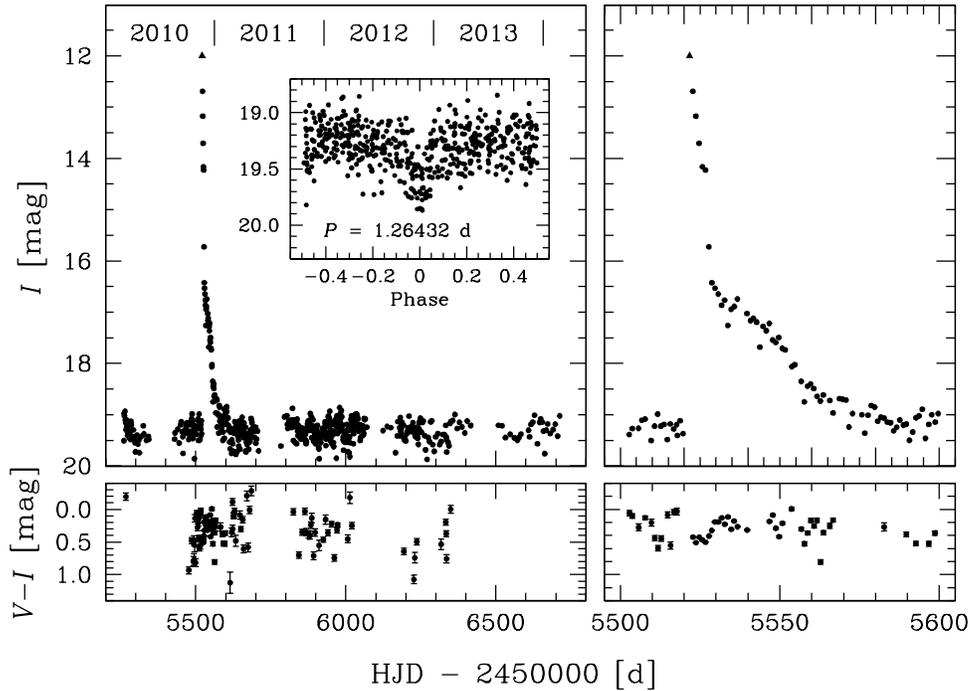}
 \caption{OGLE-IV light and colour curves of Nova LMC 1990b. The 2010 eruption is presented in the right panel. The inset shows the quiescent light curve phased with the orbital period $P_{\rm{orb}}=1.26432$~d. The triangle indicates the overexposed image.}
 \label{fig3}
\end{figure*}

We measured a period of $1.26432(8)$ d. The phased light curve (inset panel in Fig. \ref{fig3}) shows some scatter, but the eclipse-like shape is clear (U Sco and V394 CrA have similar light curves; Schaefer 2010). The $I$-band depth is 0.6 mag, while in the $V$ band it is larger, $\sim$1.5 mag. In principle, the orbital period might be two times longer, but we do not observe any differences between odd and even minima. The secondary eclipse is probably too shallow to be distinguished from flickering. 

We have tried to detect any period changes caused by the 2010 explosion, but the number of observations in the first season, before the eruption, is insufficient.

There are more photometric similarities between Nova LMC 1990b and~U~Sco-like RNe. A location on the colour--magnitude diagram indicates that the secondary is a subgiant; the nova faded rapidly ($t_3\approx 5$ d; 2.6~d for~U~Sco and 5.2~d for V394~CrA). This agrees with Sekiguchi et al. (1990a) who found a number of spectral similarities.

\section[]{V5590 Sagittari = PNV J18110375-2717276}

The discovery of this nova was announced in 2012 April (Nakano et al. 2012), but there were some reports that it was already bright in 2011 April (Walter et al. 2012).

This object has been observed in the OGLE survey since 2001. Its equatorial coordinates are $(\alpha,\delta)_{\rm{J2000.0}} = (18^{\rm{h}}11^{\rm{m}}03\fs 70,\ -27^{\circ}17{'}29\farcs 4)$. We found its counterparts in the Two-Micron All-Sky Survey (2MASS) All-Sky Catalogue of Point Sources (Cutri et al. 2003) and the Wide-field Infrared Survey Explorer (WISE) All-Sky Source Catalogue (Wright et al. 2010). The infrared photometry is listed in Table \ref{IR}.

\begin{figure*}
 \includegraphics[width=0.72\textwidth]{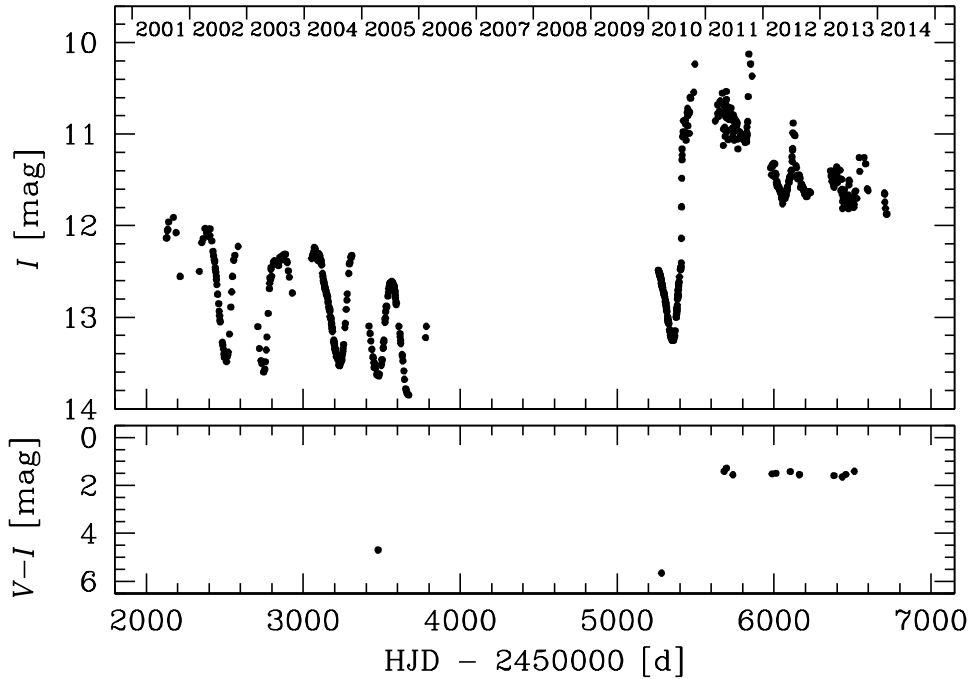}
 \caption{OGLE light and colour curves of V5590 Sgr.}
 \label{v5590}
\end{figure*}

The OGLE light curve of V5590~Sgr consists of 679 measurements in the $I$ band and 22 in the $V$ band. The data are shown in Fig. \ref{v5590}.

The eruption started on 2010 July 31 (HJD $= 2455409$) and the nova brightened by 1.5 mag in $I$ in a week (0.2 mag/d). Subsequently, the brightness rose until the end of the 2010 season (i.e. October 25) at a rate of 0.006 mag/d. The observations were resumed on 2011 March 3, when the object's brightness was the same as in 2010 October, which suggests that it had reached the maximum during a conjunction with the Sun. In the $V$ band, the amplitude of the eruption was larger than in the $I$ band, at least $\Delta V = 6$ mag.

The nova faded slowly at a rate of $0.68 \pm 0.12$ mag/yr (2011) and $0.11\pm 0.02$ mag/yr (2012-13) in $I$, and $0.99 \pm 0.68$ mag/yr (2011) and $0.16\pm 0.06$ mag/yr (2012-13) in $V$. In the $I$-band light curve, we noticed three brightenings ('jitters'; Strope et al. 2010) with amplitudes 0.5--1.0 mag lasting 20--30 d. 

The progenitor was a Mira variable with a pulsation period of $235.9 \pm 1.0$ d and an $I$-band peak-to-peak amplitude of 1.0--1.2 mag. Between 2001 and 2005, the mean brightness dropped at a constant rate of $0.19 \pm 0.01$ mag/y in $I$. The near-infrared 2MASS photometry (Cutri et al. 2003) confirmed that this object is located on sequence C (occupied by Mira stars) in the period-luminosity diagram for long period variables (Soszy\'nski et al. 2013). 

The shape of the light curve and the Mira secondary suggest that this object is a SyN of D type (i.e. a Mira variable is surrounded by an optically thick dust shell). 

\begin{table}
\centering
\begin{tabular}{ccccc}
\hline
Band & Mag & Error & Date & Source \\ \hline
$J$     & 9.294 & 0.018 & 1999-05-29 & (1) \\
$H$     & 8.127 & 0.033 & 1999-05-29 & (1) \\
$K_{s}$ & 7.349 & 0.013 & 1999-05-29 & (1) \\
$W1$    & 6.160 & 0.045 & 2010-03-21 & (2) \\
$W2$    & 5.413 & 0.024 & 2010-03-21 & (2) \\
$W3$    & 4.325 & 0.015 & 2010-03-21 & (2) \\
$W4$    & 3.650 & 0.021 & 2010-03-21 & (2) \\
\hline
\end{tabular} \\
{\scriptsize The references are (1) Cutri et al. (2003) and (2) Wright et al. (2010)}
\caption{Infrared photometry of V5590 Sgr.}
\label{IR}
\end{table}

\section[]{OGLE-2011-BLG-1444 = VVV-NOV-003}

In September 2011, the OGLE survey announced the discovery of a candidate microlensing event OGLE-2011-BLG-1444. It brightened until the end of the observing season (i.e. October 27) at a constant rate of 0.14 mag/d. However, the light curve deviated from the standard microlensing curve.

Beamin et al. (2013) have reported the discovery of the likely Galactic nova VVV-NOV-003, based on the VISTA Variables in the Via Lactea (VVV) survey data (Minniti et al. 2010). Between 2011 August and 2012 March, this object brightened by at least $\Delta K_s = 6.4$ mag, suggesting it might be a slow nova. 

\begin{figure*}
 \includegraphics[width=0.73\textwidth]{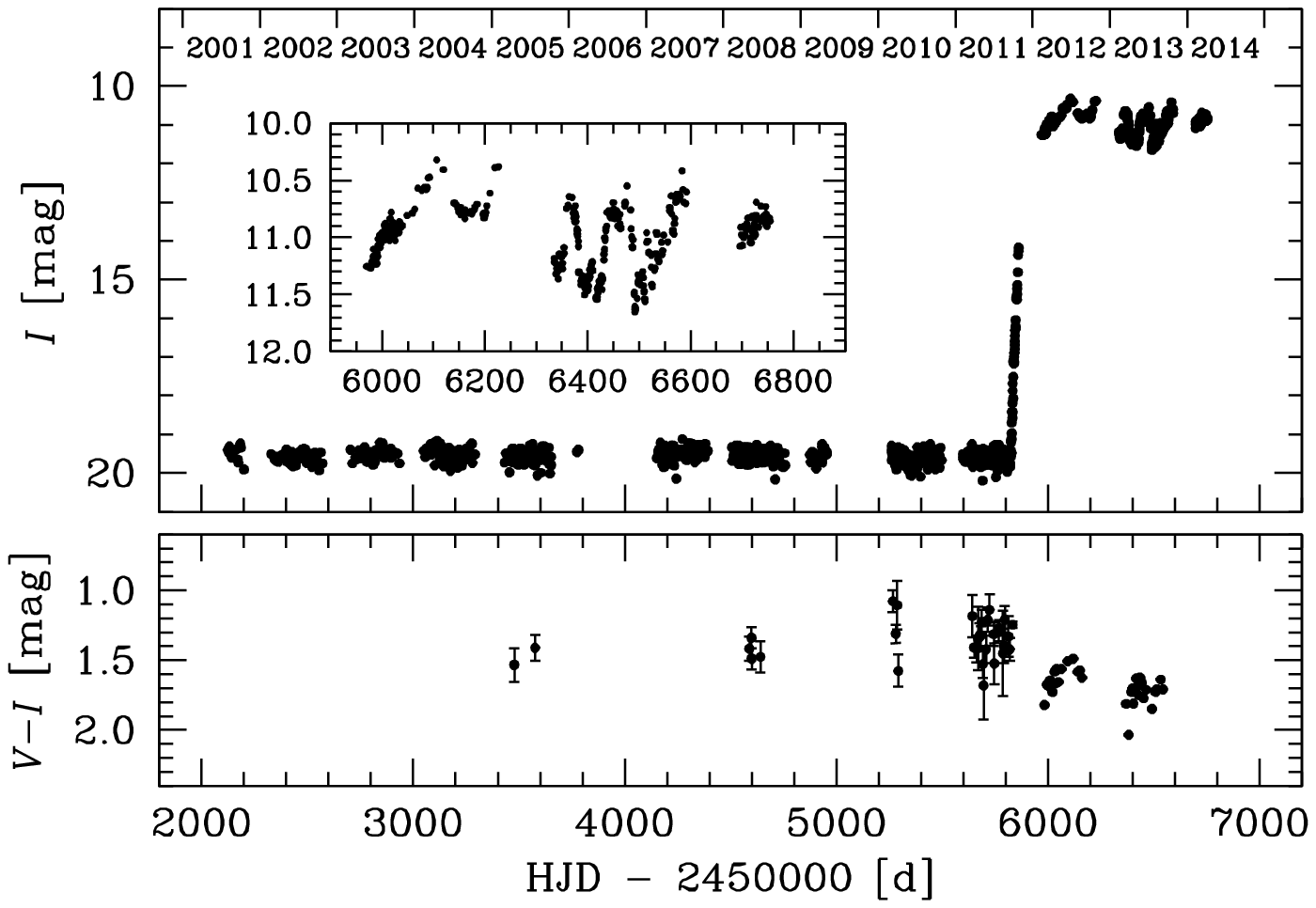}
 \caption{OGLE light and colour curves of OGLE-2011-BLG-1444. The inset panel shows a close-up of the eruption data.}
 \label{1444lc}
\end{figure*}

Poleski \& Udalski (2013) immediately realized the coincidence with OGLE-2011-BLG-1444. The OGLE data revealed an almost flat, long maximum, a distinctive feature of SyNe. The symbiotic nature of this object was later confirmed by low resolution spectroscopy (Munari et al. 2013). The light curve is presented in Fig. \ref{1444lc}.

The 2012-14 light curve (see inset panel of Fig. \ref{1444lc}) shows semiregular variations with a period of $\sim$120 d and a peak-to-peak amplitude of 1~mag. Similar variability has been observed in other SyNe (Munari et al. 2013), but its origin is unknown.

\section{Summary}

We present photometric properties of three RNe and two SyNe observed by the OGLE project. The pre-eruption light curve of V745~Sco reveals semiregular pulsations of the red giant secondary (with periods of 136.5 and 77.4 d). We do not find the ellipsoidal variability claimed by some authors. This voids previous findings based on the wrong orbital period (e.g. the distance measurement).

We report the discovery of a missing 2010 eruption of Nova LMC 1990b. Based on the eclipse-like variability in quiescence, we have measured its orbital period of $1.26432(8)$ d. The photometric properties (orbital period, colour, speed of decline) of this nova confirm it is a member of the U Sco subgroup of RNe.

We also present light curves of two likely SyNe: V5590~Sgr and OGLE-2011-BLG-1444. Both stars show flat, long maxima. The secondary of V5590~Sgr is a Mira star with a pulsation period of $235.9 \pm 1.0$ d.

We checked the OGLE photometry of old novae (almost 180 objects), located towards the Galactic bulge and the Magellanic Clouds, but none of these has shown a second eruption. We will publish these data in forthcoming papers.

Currently, the sky area of over 140 deg$^2$ is regularly monitored by the OGLE project with cadence ranging from 20~min to 2~d. Photometry is performed on-line at the telescope and transient objects are effectively found by the Early Warning System (EWS; Udalski 2003). This guarantees that only a conjunction with the Sun can cause a nova eruption to be missed. For objects brighter than $I=20$ mag, long-term photometry is available. Future (recurrent) novae eruptions located in the OGLE monitored fields can then be investigated in a more detailed way.

\section*{Acknowledgements}

We thank the anonymous referee and Dr M. Gromadzki for comments that have improved this paper. This work has been supported by the Polish Ministry of Science and Higher Education through the programme 'Ideas Plus', award No. IdP2012 000162. The OGLE project has received funding from the European Research Council under the European Community's Seventh Framework Programme (FP7/2007-2013)/ERC grant agreement No. 246678 to AU. This publication makes use of data products from WISE, which is a joint project of the University of California, Los Angeles, and the Jet Propulsion Laboratory/California Institute of Technology, funded by the National Aeronautics and Space Administration.

\bsp

\label{lastpage}

\end{document}